\documentclass[useAMS,usenatbib,usegraphicx]{mn2e}
\newcommand{\ul}{\underline{\hspace{40pt}}}
\newcommand{\cs}{c_{\rm s}}

\begin{document}

\title[Molecular Cloud Fragmentation]
{Three-dimensional simulations of molecular cloud fragmentation regulated 
by magnetic fields and ambipolar diffusion}
\author[T. Kudoh et al.]{Takahiro Kudoh,$^{1}$\thanks{E-mail:
kudoh@th.nao.ac.jp (TK)} Shantanu Basu,$^{2}$ 
Youichi Ogata$^3$ and Takashi Yabe$^4$ \\
$^{1}$ Division of Theoretical Astronomy, National Astronomical Observatory of Japan,
Mitaka, Tokyo 181-8588, Japan \\
$^{2}$Department of Physics and Astronomy, University of Western Ontario, 
London, Ontario, N6A 3K7, Canada \\
$^{3}$Department of Mechanical Engineering and Science, Tokyo Institute of Technology, 
O-okayama, Meguro-ku, Tokyo 152-8552, Japan \\
$^{4}$Integrated Research Institute, Solutions Research Organization, 
Tokyo Institute of Technology, O-okayama, Meguro-ku, Tokyo 152-8552, Japan \\
}

\date{Accepted 2007 June 14}

\maketitle

\label{firstpage}

\begin{abstract}
We employ the first fully three-dimensional simulation to study the
role of magnetic fields and ion-neutral friction in regulating 
gravitationally-driven fragmentation of molecular clouds. 
The cores in an initially subcritical cloud develop gradually over an 
ambipolar diffusion time
while the cores in an initially supercritical cloud develop in a 
dynamical time. 
The infall speeds on to cores are subsonic in the case of 
an initially subcritical cloud, while an extended ($\ga 0.1$ pc) region
of supersonic infall exists in the case of an initially supercritical cloud.
These results are consistent with previous two-dimensional simulations.
We also found that a snapshot of the relation between 
density ($\rho$) and the strength of the magnetic field ($B$) 
at different spatial points of the cloud coincides with 
the evolutionary track of an individual core. 
When the density becomes large, both relations 
tend to $B \propto \rho^{0.5}$. 

\end{abstract}


\section{Introduction}

Magnetic fields in molecular clouds play an important role in
the early stages of star formation. They may regulate the
cloud fragmentation process, moderate the infall motions on to
density peaks, control angular momentum evolution through magnetic braking, 
launch jets from the near-protostellar environment, and possibly
determine a finite mass reservoir for star formation by 
limiting accretion from a magnetically-dominated envelope.
The prevailing macroturbulence in molecular cloud 
envelopes also likely represents magnetohydrodynamic motions.
This paper concerns itself with the first two issues above: 
we employ the first fully three-dimensional simulation to study the
role of magnetic fields {\it and} ion-neutral friction 
in regulating gravitationally-driven fragmentation and infall. 

The relative strengths of gravity and the magnetic field can be 
quantified through the mass-to-flux ratio $M/\Phi$. There exists
a critical mass-to-flux ratio $(M/\Phi)_{\rm crit}$ 
\citep{mes56,str66,mou76,tom88} such that if 
$M/\Phi > (M/\Phi)_{\rm crit}$, a pressure-bounded cloud
is supercritical and is prone to indefinite collapse 
if the external pressure exceeds some value. This behaviour is
analogous to that of the nonmagnetic Bonnor-Ebert sphere. Conversely, if 
$M/\Phi < (M/\Phi)_{\rm crit}$, a cloud is subcritical and cannot
collapse even in the limit of infinite external pressure, as long as
magnetic flux-freezing applies. A similar condition
$M/\Phi < (M/\Phi)_{\rm crit} = (2 \pi G^{1/2})^{-1}$ is required for
unconditional stability of 
an infinite uniform (in $x,y$) layer that is flattened along the $z$-direction
of a background magnetic field \citep{nak78}. The various numerical 
values of $(M/\Phi)_{\rm crit}$
differ by small factors of order unity, and we adopt the result of
\citet{nak78} since their model closely resembles the initial state in
our calculation. 

Magnetic field strength measurements through the Zeeman effect 
reveal that the mass-to-flux ratios are clustered about the 
critical value for collapse \citep{cru99,cru04,shu99} and that 
there is also an approximate equipartition
between the absolute values of gravitational energy and nonthermal
(hereafter, ``turbulent'') energy \citep{mye88,bas00}.
Measurements of polarized emission from dust grains, which reveal 
the field morphology, generally indicate that the field in cloud
cores is well ordered and not dominated by turbulent motions,
with application of the Chandrasekhar-Fermi method yielding
mass-to-flux ratios near the critical value
\citep{lai01,lai02,cru04b,cur04,kir06}.

\citet{mes56} pointed out that  
even if clouds are magnetically supported, 
ambipolar diffusion (resulting from ion-neutral slip) will cause
the support to be lost and stars to form. More specifically,
a medium with a subcritical mass-to-flux ratio will still 
undergo a gravitationally-driven instability, occurring
on the ambipolar diffusion time scale rather than
the dynamical time scale \citep{lan78,zwe98,cio06}. The 
length scale of the instability is essentially the Jeans
scale in the limit of highly subcritical clouds (the same
length scale as for highly supercritical fragmentation) but
can be much larger when the mass-to-flux ratio is
close to the critical value \citep{cio06}. 
Most nonlinear calculations of ambipolar diffusion
driven evolution have focused on a single axisymmetric 
core, but newer models focus on a fragmentation process
that results in the formation of multiple cores and somewhat
irregular density and velocity structure. 
\citet{ind00} carried out a two-dimensional simulation
of an infinitesimally-thin sheet threaded by an initially 
perpendicular magnetic field.
Starting with slightly subcritical initial conditions, they 
followed the initial growth of mildly elongated fragments which 
occurred on a time scale intermediate between the dynamical
time associated with supercritical collapse and the ambipolar diffusion
time-scale associated with highly subcritical clouds. 
\citet{bas04} carried out two-dimensional simulations of 
a magnetized sheet in the thin-disc approximation, which 
incorporates a finite disc half-thickness $Z$
consistent with hydrostatic equilibrium and thereby
includes the effect of magnetic pressure.
They studied a model which had an initially critical
mass-to-flux ratio and another which was supercritical
by a factor of two. One of their main results was that
the critical model had subsonic (maximum speed $\approx
0.5 \cs$, where $\cs$ is the isothermal sound speed) infall,
while the decidedly supercritical cloud had infall speeds
$\ga 1 \cs$ on scales $\sim 0.1$ pc from the core centres.
This is a significant 
observationally-testable difference between dynamical
(supercritical) fragmentation and ambipolar-diffusion regulated
(critical or subcritical) fragmentation. Yet another mode of
fragmentation is the so-called turbulent fragmentation, which in
fact corresponds to collapse driven by a strong external compression.
\citet{li04} and \citet{nak05} have studied this process for a 
magnetized sheet using the thin-disc approximation, 
and including the effect of ion-neutral friction. They find that 
a mildly subcritical cloud can undergo locally rapid ambipolar 
diffusion and form multiple fragments because of an initial 
large-scale highly supersonic compression wave. The core formation
occurs on a crossing time of the simulation box, which is related
to the dynamical time. \citet{li04} estimate that such a process can 
simultaneously maintain a relatively low efficiency of star formation,
as is required by observations \citep{lad03}.

In this paper, we study the three-dimensional extension of models
such as those of \citet{ind00} and \citet{bas04}. The self-consistent
calculation of the vertical structure of the cloud allows us to 
test the predictions of two-dimensional models as well as to make 
some new predictions. We model clouds
that are either decidedly supercritical or subcritical and study the 
evolution after the introduction of small-amplitude perturbations.
The case of compression-induced collapse will be studied in a
separate paper. We note that our three-dimensional model is not
a cubic region but rather a flattened three-dimensional layer that
is consistent with the expected settling of gas along the
direction of the magnetic field. In reality, we believe that our modeled
region would represent the dense midplane of a larger more turbulent
cloud. As demonstrated by the one-dimensional models of 
\citet{kud03,kud06}, the turbulent motions in a stratified magnetized 
cloud develop the largest amplitude motions in the outer low-density
envelope, while maintaining transonic or subsonic motions near the
midplane. We believe that the fragmentation process as modeled 
in this paper may proceed while long-lived turbulent motions continue
on larger scales.

\section{Numerical Model}

\subsection{Basic Equations}
\label{equation}

We solve the three-dimensional magnetohydrodynamic 
(MHD) equations including self-gravity and ambipolar diffusion,
assuming that neutrals are much more numerous than ions:

\begin{equation}
\frac{\partial \rho}{\partial t} 
+ \mbox{\boldmath$v$} \cdot \nabla \rho
= -\rho \nabla \cdot \mbox{\boldmath$v$},
\label{eq:continuity}
\end{equation}

\begin{equation}
\frac{\partial \mbox{\boldmath$v$}}{\partial t} 
+ (\mbox{\boldmath$v$} \cdot \nabla) \mbox{\boldmath$v$}
=-\frac{1}{\rho} \nabla p
+ \frac{1}{c \rho} \mbox{\boldmath$j$} \times \mbox{\boldmath$B$}
- \nabla \psi,
\label{eq:momentum}
\end{equation}

\begin{equation}
\frac{\partial \mbox{\boldmath$B$}}{\partial t}
= \nabla \times (\mbox{\boldmath$v$} \times \mbox{\boldmath$B$})
+ \nabla \times \left[\frac{\tau_{ni}}{c\rho} (\mbox{\boldmath$j$} \times \mbox{\boldmath$B$}) \times \mbox{\boldmath$B$}\right],
\label{eq:induction}
\end{equation}

\begin{equation}
\mbox{\boldmath$j$}=\frac{c}{4\pi} \nabla \times \mbox{\boldmath$B$},
\end{equation}

\begin{equation}
\nabla ^2 \psi = 4\pi G \rho,
\label{eq:poisson}
\end{equation}

\begin{equation}
p=c_s^2 \rho,
\end{equation}
where $\rho$ is the density of neutral gas, $p$ is the pressure, 
{\boldmath$v$} is the velocity, {\boldmath$B$} is the magnetic field,
{\boldmath$j$} is the electric current density, 
$\psi$ is the self-gravitating potential, 
and $c_s$ is the sound speed.
Instead of solving a detailed energy equation, we assume isothermality
for each Lagrangian fluid particle 
\citep{kud03,kud06}:
\begin{equation}
\frac{d c_s}{dt} 
= \frac{\partial c_s}{\partial t} 
+ \mbox{\boldmath$v$} \cdot \nabla c_s = 0.
\label{eq:isothermal}
\end{equation}
For the neutral-ion collision time in equation (\ref{eq:induction})
and associated quantities,
we follow \citet{bas94}, so that
\begin{equation}
\tau_{ni} =  1.4\, \frac{m_i+m_n}{\rho_i \langle \sigma w \rangle_{in}},
\label{eq:tau}
\end{equation}
where $\rho_i$ is the density of ions and $\langle\sigma w\rangle_{in}$ 
is the average collisional rate between ions of mass $m_i$ and neutrals
of mass $m_n$. Here, we use typical values of HCO$^+$-H$_2$ collisions, 
for which $\langle\sigma w\rangle_{in}=1.69 \times 10^{-9}$ cm$^{-3}$s$^{-1}$ 
and $m_i/m_n =14.4$. We also assume that
the ion density $\rho_i$ is determined by the approximate relation
\citep{elm79,nak79}
\begin{equation}
\rho_i=m_i K \left(\frac{\rho/m_n}{10^5 \mbox{cm$^{-3}$}}\right)^k,
\end{equation}
where we assume $K=3 \times 10^{-3}$cm$^{-3}$ and $k=0.5$
throughout this paper.

\subsection{Initial Conditions}

As an initial condition, we assume hydrostatic equilibrium of 
a self-gravitating one-dimensional cloud along $z$-direction
\citep{kud03,kud06}. The hydrostatic equilibrium is calculated 
from equations

\begin{equation}
\frac{dp}{dz}=\rho g_z,
\label{eq:hsp}
\end{equation}

\begin{equation}
\frac{dg_z}{dz}=-4\pi G \rho,
\label{eq:hsg}
\end{equation}

\begin{equation}
p=c_s^2 \rho,
\label{eq:hss}
\end{equation}
subject to the boundary conditions
\begin{equation}
g_z(z=0)=0,\ \ \rho(z=0)=\rho_0,\ \ p(z=0)=\rho_0 c_{s0}^2
\end{equation}
where $\rho_0$ and $c_{s0}$ are the initial density and
sound speed at $z=0$.
If the initial sound speed (temperature) is uniform throughout the region,
we have the following analytic solution $\rho_S$ found by \citet{spi42}:
\begin{equation}
\rho_S(z)=\rho_0 \,\mbox{sech}^2(z/H_0),
\end{equation}
where
\begin{equation}
H_0=\frac{c_{s0}}{\sqrt{2\pi G \rho_0}}
\end{equation}
is the scale height.
However, an isothermal molecular cloud is usually surrounded
by warm material, such as neutral hydrogen gas.
Hence, we assume the initial sound speed distribution to be
\begin{equation}
c_{s}^2(z)=c_{s0}^2 
+ \frac{1}{2} (c_{sc}^2 - c_{s0}^2)
\left[ 1+\tanh \left(\frac{|z|-z_c}{z_d}\right) \right]
\end{equation}
where we take $c_{sc}^2=10\,c_{s0}^2$, $z_c=2H_0$, and $z_d=0.1H_0$
throughout the paper. By using this sound speed distribution, we can solve equations
(\ref{eq:hsp})-(\ref{eq:hss}) numerically. The initial density distribution 
of the numerical solution shows that it is almost the same as Spitzer's 
solution for $0 \leq z \leq z_c$.

We also assume that the initial magnetic field is uniform along the $z$-direction:
\begin{equation}
B_z=B_0,\ \ B_x=B_y=0,
\end{equation}
where $B_0$ is constant.

In this equilibrium sheet-like gas, we input a random velocity perturbation
\citep{miy87b} at each grid point:
\begin{equation}
v_x=0.1c_{s0}R_m,\ \ v_y=0.1c_{s0}R_m,\ \ v_z=0.0
\end{equation}
where $R_m$ is a random number chosen uniformly from the range [-1,1].
The $R_m$'s for each of $v_x$ and $v_y$ are independent realizations.
However, each model presented in this paper uses the same pair of realizations
of $R_m$ for generating the initial perturbations.

\subsection{Numerical Parameters}

A set of fundamental units for this problem are
$c_{s0}$, $H_0$, and $\rho_0$. These yield a time 
unit $t_0=H_0/c_{s0}$. The initial magnetic field
strength introduces one dimensionless free parameter,
\begin{equation}
\beta_0 \equiv \frac{8 \pi p_0}{B_0^2} 
= \frac{8 \pi \rho_0 c_{s0}^2}{B_0^2},
\end{equation}
the ratio of gas to magnetic pressure 
at $z=0$.

In the sheet-like equilibrium cloud with a vertical 
magnetic field, $\beta_0$ is related to the mass-to-flux ratio
for Spitzer's self-gravitating cloud. The mass-to-flux ratio 
normalized to the critical value is 
\begin{equation}
\mu_S \equiv 2\pi G^{1/2} \frac{\Sigma_S}{B_0} 
\end{equation}
where
\begin{equation}
\Sigma_S=\int_{-\infty}^{\infty} \rho_S dz = 2\rho_0 H_0
\end{equation}
is the column density of Spitzer's self-gravitating cloud.
Therefore,
\begin{equation}
\beta_0=\mu_S^2.
\end{equation}
Although the initial cloud we used is not exactly 
the same as the Spitzer cloud, $\beta_0$ is a good 
indicator to whether or not the magnetic field 
can prevent gravitational instability
\citep{nak78}.

Dimensional values of all quantities can be found 
through a choice of $\rho_0$ and $c_{s0}$.
For example, for $c_{s0}=0.2$ km s$^{-1}$ and $n_0=\rho_0/m_n=10^4$ cm$^{-3}$, 
we get $H_0=0.05$ pc, $t_0=2.5 \times 10^5$ yr, and 
$B_0=20\,\mu$G if $\beta_0=1$.

\subsection{Numerical Technique}

In order to solve the equations numerically, we use the 
CIP method \citep{yab91a,yab01}
for equations (\ref{eq:continuity}), (\ref{eq:momentum}) 
and (\ref{eq:isothermal}), and the method of 
characteristics-constrained transport (MOCCT; \citet{sto92}) 
for equation (\ref{eq:induction}), including an explicit integration
of the ambipolar diffusion term. The combination of the CIP and 
MOCCT methods is summarized in \citet{kud99} and
\citet{oga04}. It includes the CCUP method 
\citep{yab91b} for the calculation of gas pressure, 
in order to get more numerically stable results.
The numerical code in this paper is based on 
that of \citet{oga04}. 

In this paper, the ambipolar diffusion term is only included when the
density is greater than a certain value, $\rho_{cr}$.
We let $\rho_{cr}=0.3\rho_0$ both for numerical convenience
and due to the physical idea 
that the low density region is affected 
by external ultraviolet radiation so that the ionization fraction
becomes large, i.e. $\tau_{ni}$ becomes small
\citep{cio95}.
Under this assumption, the upper atmosphere of the sheet-like cloud
is not affected by ambipolar diffusion.
This simple assumption helps to avoid very small time-steps
due to the low density region in order to maintain stability of
the explicit numerical scheme. 

We used a mirror-symmetric boundary 
condition at $z=0$ and periodic boundaries in the $x$ 
and $y$-directions. At the upper boundary 
at $z=z_{\rm out}=4H_0$, we also used a mirror-symmetric boundary 
except when we solve the gravitational potential. This symmetric condition 
is just for numerical convenience. However, because 
the results we show later in this paper are consistent with previous 
two-dimensional simulations, we believe that the boundary conditions 
do not affect the result significantly. The Poisson equation (\ref{eq:poisson}) 
is solved by the Greens function method to compute the gravitational 
kernels in $z$-direction, along with a Fourier transform 
method in the $x$ and $y$-directions \citep{miy87b}.
This method of solving the Poisson equation allows us to 
find the gravitational potential of a vertically isolated cloud 
within $|z|<z_{\rm out}$.

The computational region is $|x|,|y| \leq 8\pi H_0$ and 
$0 \leq z \leq 4H_0$. 
The number of grid points for each direction is $(N_x,N_y,N_z)=(64,64,40)$.
Since the most unstable wavelength for no magnetic field is about 
$4\pi H_0$ 
\citep{miy87a},
we have 16 grid points within this wavelength.
We have also 10 grid points within the scale height of the initial cloud
in the $z$-direction. While this is not a high-resolution simulation,
we believe that we have the minimum number of grid points 
to study the gravitational instability, especially by using 
the code based on CIP \citep{oga04}.
The maximum computational time, which occurs for the case of the 
subcritical cloud,
is about 85 hours of CPU time using a single processor 
of the VPP5000 in the National Astronomical Observatory of Japan.

\section{Results}
\label{results}

\begin{figure}
  \resizebox{\hsize}{!}{\includegraphics{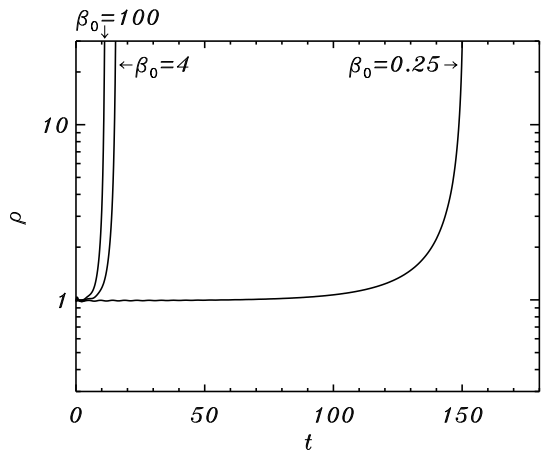}}
      \caption{
The time evolution of the density at the location 
where the density reaches its maximum value in each simulation.
Each line shows a case of different $\beta_0$. 
      }
         \label{fig1}
\end{figure}
Figure 1 shows the time evolution of the density at the location 
where the density reaches its maximum value in each simulation.
The simulations are stopped when the maximum density is about
$30 \rho_0$.
Each line shows a case of different 
$\beta_0$. When $\beta_0$ is 100 or 4, the magnetic field is not 
strong enough to suppress the self-gravitational instability of 
the cloud. In these cases, the density evolves rapidly, over the sound-crossing 
time of the most unstable wavelength ($\sim 4\pi H_0$). 
However, when $\beta_0=0.25$, the cloud is self-gravitationally 
stable unless ion-neutral slip is present. Therefore, the density evolves 
gradually over the diffusion time of the magnetic field.
According to the two-dimensional linear analysis by \citet{cio06},
the evolutionary time scale of a significantly subcritical cloud is
about ten times longer than the dynamical time,
for a standard ionization fraction, as used here.
Our numerical result is consistent with their analysis.

\begin{figure}
  \resizebox{\hsize}{!}{\includegraphics{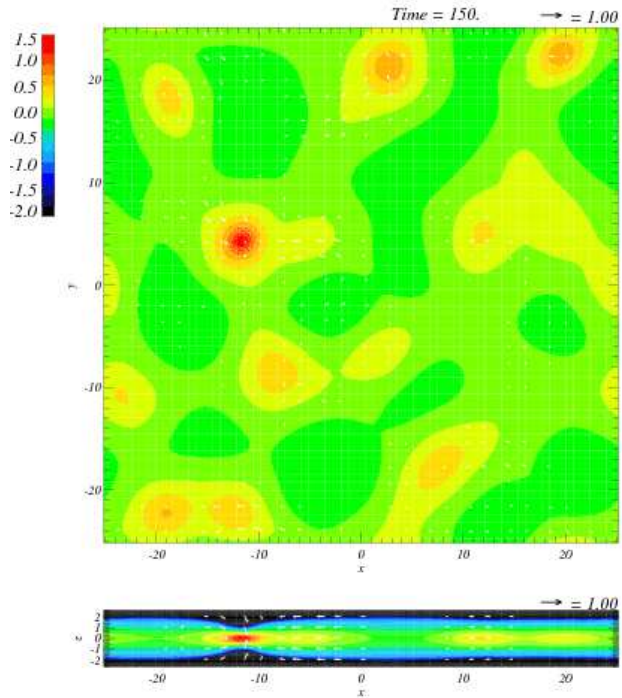}}
      \caption{
      The logarithmic density contours for $\beta_0=0.25$ at $t=150$.
      Arrows show velocity vectors on each cross section. 
      Upper panel shows the cross section at $z=0$, and the lower panel 
      shows the cross section at $y=4.3$.
      }
         \label{fig2}
\end{figure}
\begin{figure}
  \resizebox{\hsize}{!}{\includegraphics{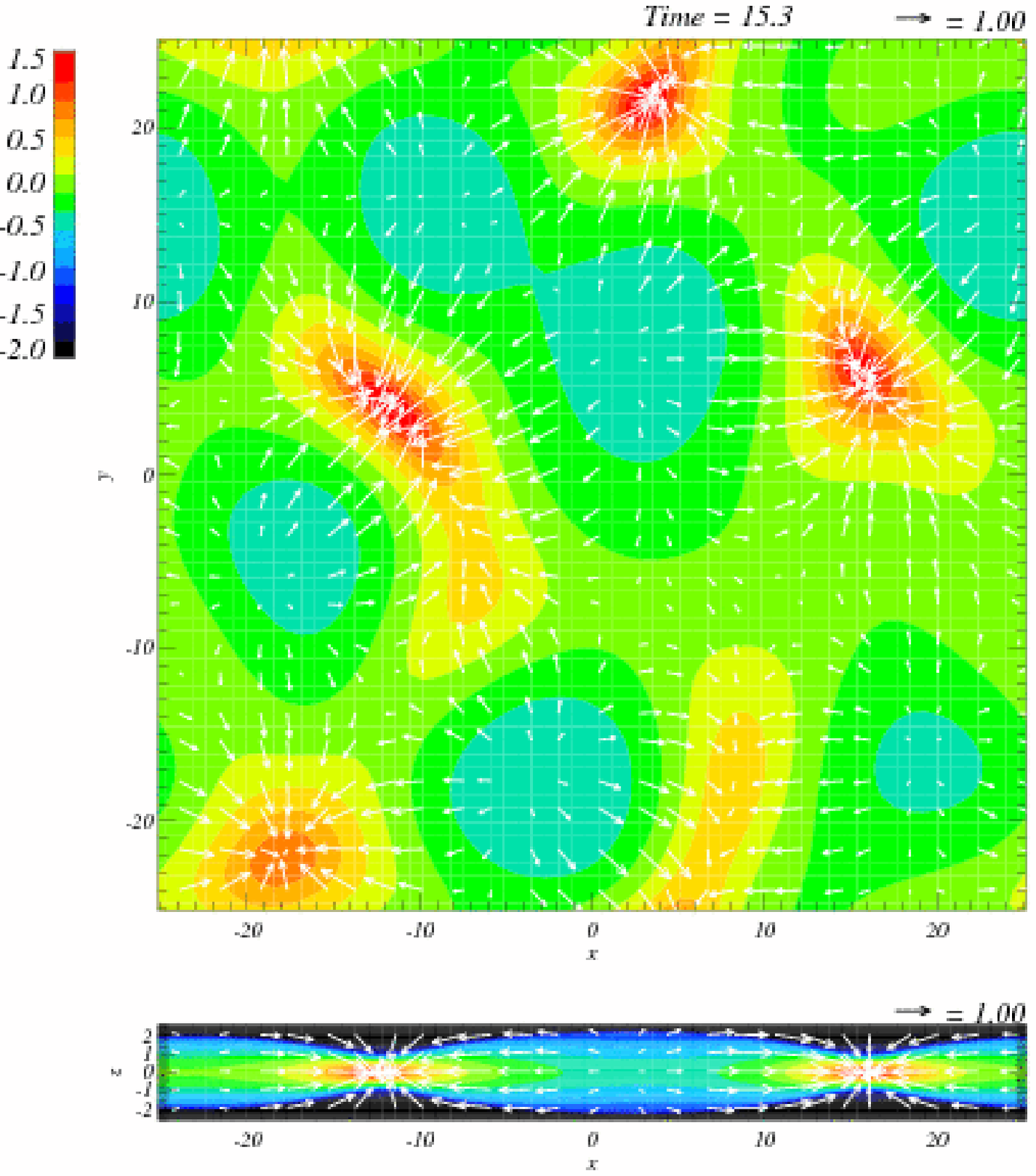}}
      \caption{
      The logarithmic density contours for $\beta_0=4$ at $t=15.3$.
      Arrows show velocity vectors on each cross section. 
      Upper panel shows the cross section at $z=0$, and the lower panel 
      shows the cross section at $y=5.1$.
      }
         \label{fig3}
\end{figure}
\begin{figure}
  \resizebox{\hsize}{!}{\includegraphics{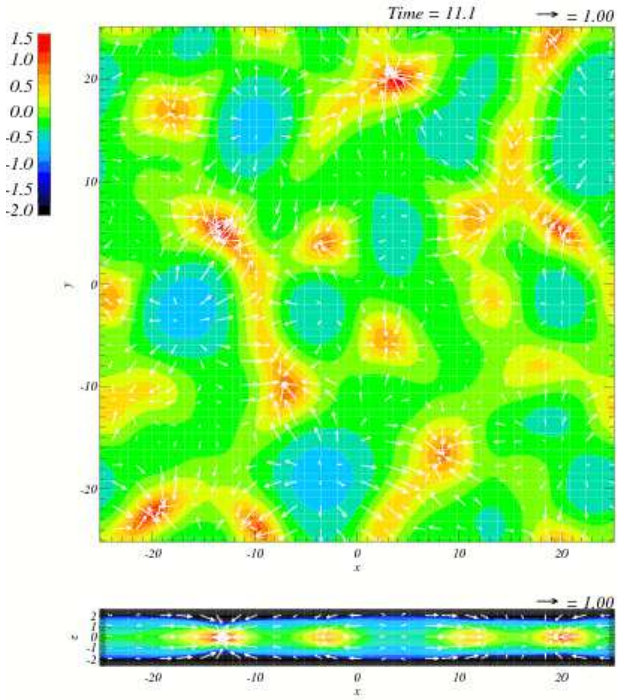}}
      \caption{
      The logarithmic density contours for $\beta_0=100$ at $t=11.1$.
      Arrows show velocity vectors on each cross section. 
      Upper panel shows the cross section at $z=0$, and the lower panel 
      shows the cross section at $y=5.1$.
      }
         \label{fig4}
\end{figure}
Figure 2, Figure 3, and Figure 4 show the logarithmic density 
contours for $\beta_0=0.25$ at $t=150$, $\beta_0=4$ at $t=15.3$, 
and $\beta_0=100$ at $t=11.1$ respectively. Each upper panel shows 
the cross section at $z=0$, and the lower panel shows the cross 
section at $y=4.3$ for $\beta_0=0.25$, $y=5.1$ for $\beta_0=4$, 
and $y=5.1$ for $\beta_0=100$ respectively. 
The values of $y$ for the lower panels are chosen so that the 
vertical cut passes through at least one dense core.
(In these numerical simulations, we use the term "core"
to refer to the region where the density is greater than the mean
background density by about a factor of 3.)
The size of cores for $\beta_0=4$ is bigger 
than that of $\beta_0=100$.  The size becomes smaller again 
when the magnetic field is stronger than critical ($\beta_0=0.25$). 
This result is consistent with the two-dimensional linear analysis 
of \citet{cio06}, who found a hybrid mode for critical or 
mildly supercritical clouds in which the combined effect of field-line
dragging and magnetic restoring forces enforce a larger than usual
fragmentation scale.
Arrows show velocity vectors on each cross section. 
Maximum velocities become supersonic for $\beta_0=4$ and $\beta_0=100$, 
but remain subsonic for $\beta_0=0.25$. This is also consistent 
with the two-dimensional numerical simulations of \citet{bas04}.

\begin{figure}
  \resizebox{\hsize}{!}{\includegraphics{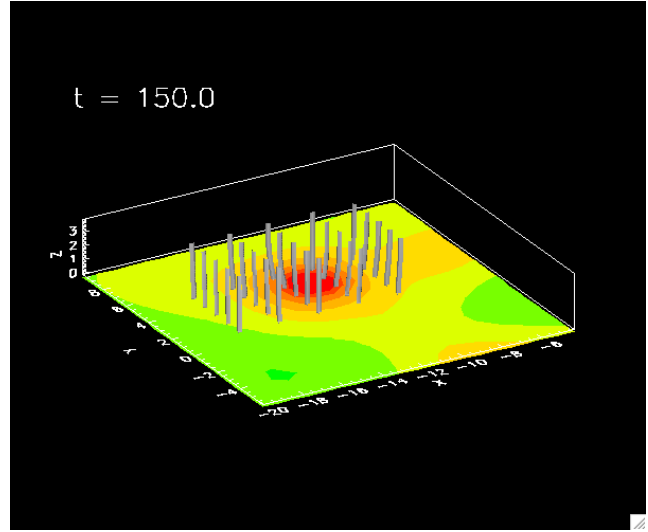}}
      \caption{
The close up views of the density contours around cores for $\beta_0=0.25$.
Magnetic field lines near cores are also plotted in three-dimensional space.
      }
         \label{fig5}
\end{figure}
\begin{figure}
  \resizebox{\hsize}{!}{\includegraphics{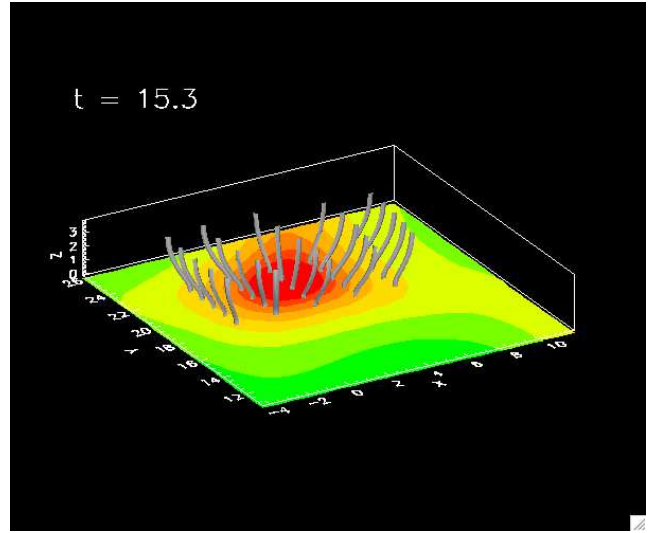}}
      \caption{
The close up views of the density contours around cores for $\beta_0=4$.
Magnetic field lines near cores are also plotted in three-dimensional space.
      }
         \label{fig6}
\end{figure}
Figure 5 and Figure 6 shows the close up views of the density 
contours around cores for $\beta_0=0.25$ and $\beta_0=4$ respectively. 
Magnetic field lines near cores are also plotted in three-dimensional space. 
When $\beta_0=0.25$, the neutral gas has to slip through the field lines 
to make a gravitationally bound core. Therefore, the field lines are 
not so deformed in the case of $\beta_0=0.25$, in contrast to 
those of $\beta_0=4$.

\begin{figure}
  \resizebox{\hsize}{!}{\includegraphics{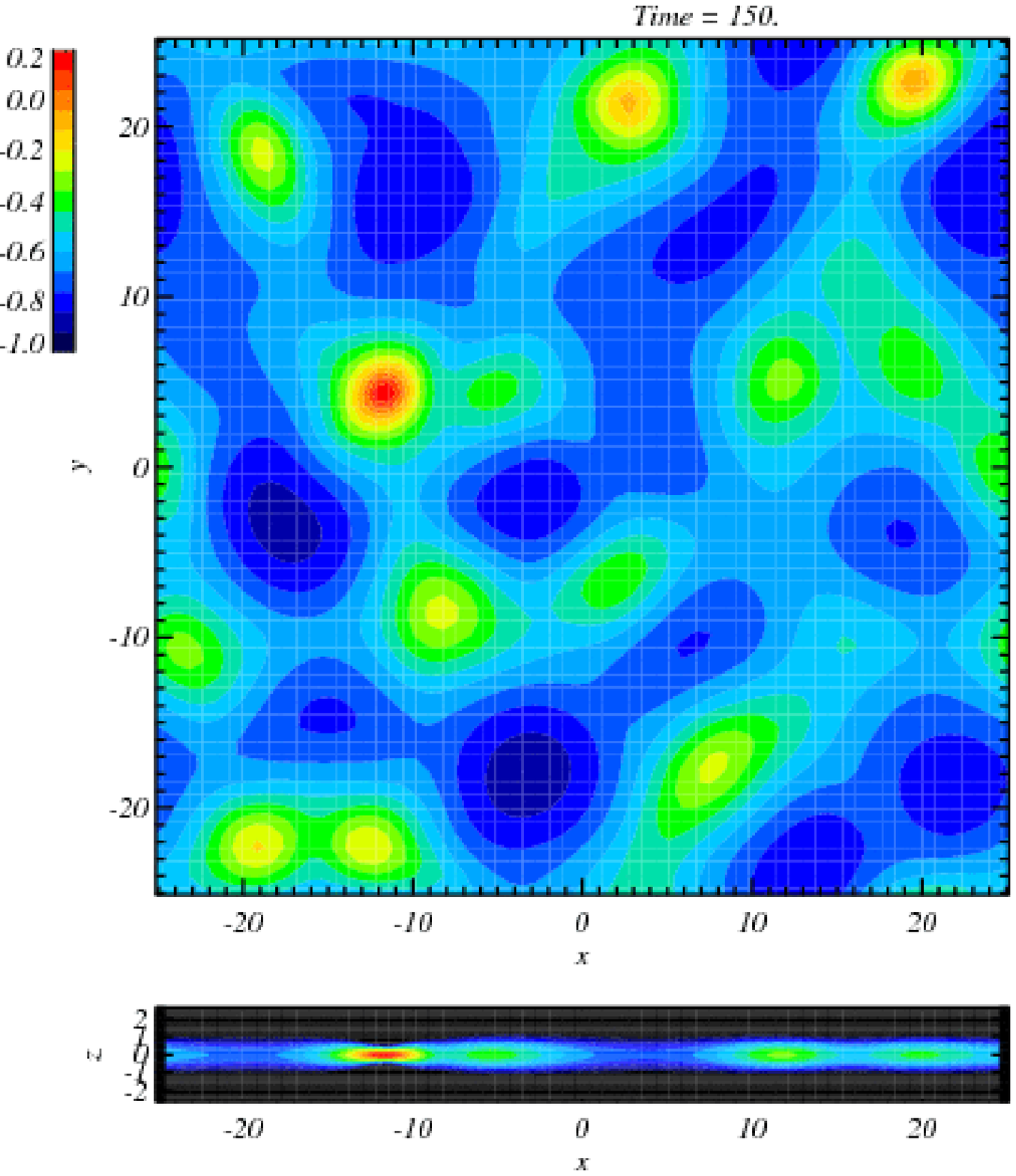}}
      \caption{
The logarithmic contours of plasma $\beta$ for $\beta_0=0.25$ at $t=150$.
Upper panel shows the cross section at $z=0$, and the lower panel 
shows the cross section at $y=4.3$.
      }
         \label{fig7}
\end{figure}
\begin{figure}
  \resizebox{\hsize}{!}{\includegraphics{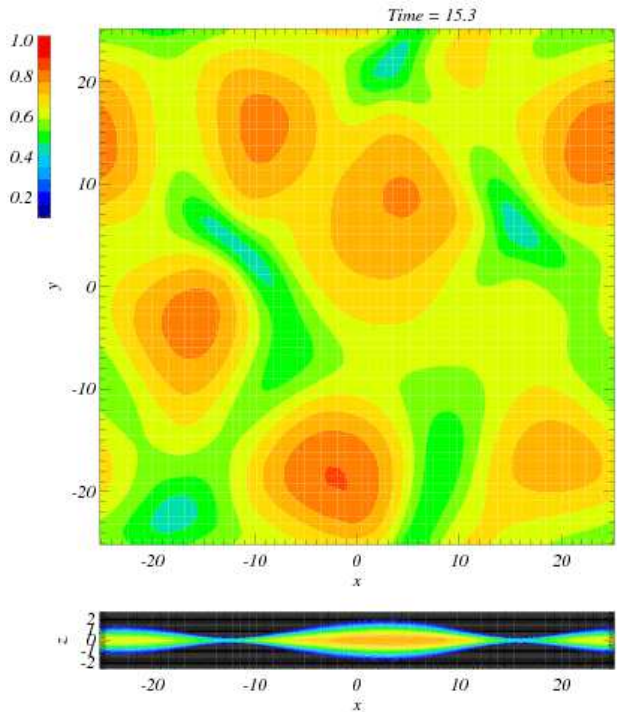}}
      \caption{
The logarithmic contours of plasma $\beta$ for $\beta_0=4$ at $t=15.3$.
Upper panel shows the cross section at $z=0$, and the lower panel 
shows the cross section at $y=5.1$.
      }
         \label{fig8}
\end{figure}
Figure 7 and Figure 8 show the logarithmic contours of 
plasma $\beta$ for $\beta_0=0.25$ at $t=150$ and $\beta_0=4$ 
at $t=15.3$, respectively. When $\beta_0=0.25$, 
the plasma $\beta$ in cores is {\it greater than in the surroundings}. 
This is because the mass-to-flux ratio (and therefore $\beta$)
has to increase in order for the core to become gravitationally
unstable.
The maximum $\beta$ is larger than 1 at the centre of a core, 
which means that the centre of the core is approximately supercritical. 
In contrast to this, the plasma $\beta$ in cores is {\it slightly lower 
than the surroundings} when $\beta_0=4$. In this case, 
the magnetic field is swept up by the contracting cloud 
before ion-neutral slip works efficiently. 
If hydrostatic equilibrium along the $z$-direction is exactly satisfied, 
the plasma $\beta$ would remain constant in time and space. 
The slightly lower values of $\beta$ in cores are probably caused 
by the modestly nonequilibrium state along $z$ during the evolution.

\begin{figure}
  \resizebox{\hsize}{!}{\includegraphics{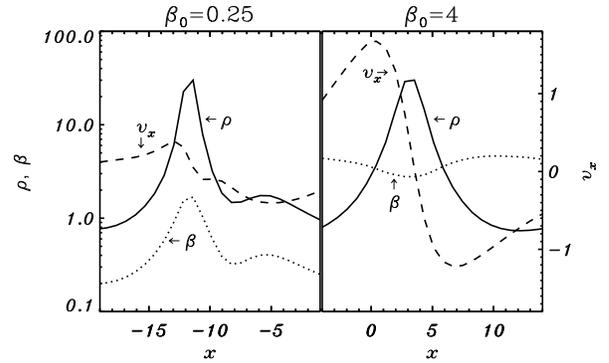}}
      \caption{
The densities, plasma $\beta$, and $v_x$ along $x$-axes 
for lines that cut through the cores shown in Figure 5 and Figure 6. 
The left panel shows the core for $\beta_0=0.25$. 
The right panel shows the core for $\beta_0=4$. 
      }
         \label{fig9}
\end{figure}
\begin{figure}
  \resizebox{\hsize}{!}{\includegraphics{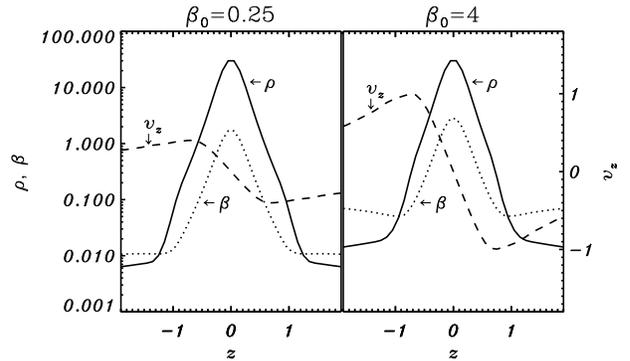}}
      \caption{
The densities, plasma $\beta$, and $v_z$ along $z$-axes 
for lines that cut through the cores shown in Figure 5 and Figure 6. 
The left panel shows the core for $\beta_0=0.25$. 
The right panel shows the core for $\beta_0=4$. 
      }
         \label{fig10}
\end{figure}
Figure 9 shows the densities, plasma $\beta$, and $v_x$
along $x$-axes for lines that cut through the cores 
shown in Figure 5 and Figure 6. The left panel shows 
the core for $\beta_0=0.25$. The right panel shows the core 
for $\beta_0=4$. Figure 9 shows that the plasma $\beta$ in 
the core is higher than the surroundings for $\beta_0=0.25$ and
lower for $\beta_0=4$. The infall velocity 
for $\beta_0=4$ shows supersonic values, while the velocity for 
$\beta_0=0.25$ is subsonic. These velocities are comparable to 
those in Figure 2 and Figure 3 of \citet{bas04}.  Figure 10 shows 
the densities, plasma $\beta$, and $v_z$ along $z$-axes 
for lines that cut through the same cores. It also shows that 
the infall velocity for $\beta_0=4$ reaches mildly supersonic values, 
while the velocity for $\beta_0=0.25$ is subsonic. 

\begin{figure}
  \resizebox{\hsize}{!}{\includegraphics{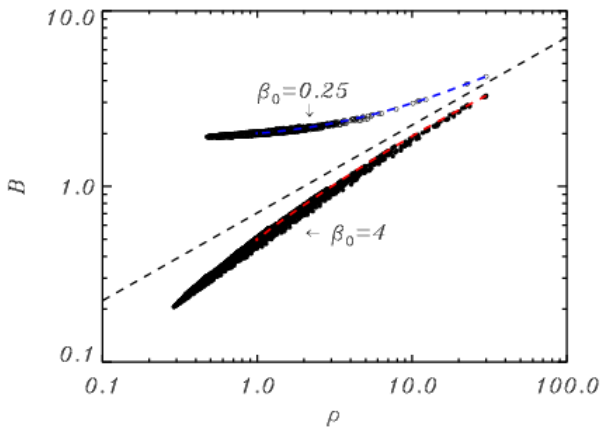}}
      \caption{
Open circles show the magnetic field strength as a function of density 
along $z=0$ at $t=150$ for the model with $\beta_0=0.25$. 
The strength of the magnetic field is normalized
by $\sqrt{8\pi \rho_0 c_{s0}^2}$. 
Filled circles are the same for $\beta_0=4$, at $t=15.3$. 
The blue line shows the evolutionary track of the point at which the density 
achieves its maximum value for the model with $\beta_0=0.25$. 
The red line is the same for $\beta_0=4$.
      }
         \label{fig11}
\end{figure}
Figure 11 shows the relation between density and magnetic field 
on the plane $z=0$. The strength of the magnetic field is normalized
by $\sqrt{8\pi \rho_0 c_{s0}^2}$. Open circles show the magnetic field strength 
as a function of density along $z=0$ at $t=150$ for the model with
$\beta_0=0.25$ (see Fig. 2). Filled circles are the same 
for $\beta_0=4$, at $t=15.3$ (see Fig. 3). The blue line 
shows the evolutionary track of the point at which the density 
achieves its maximum value for the model with $\beta_0=0.25$. 
The red line is the same but for $\beta_0=4$. This figure 
shows that {\it the snapshot of the relation between density and 
magnetic field at different spatial points in the midplane of the cloud overlaps 
with the evolutionary track of an individual core} . The dashed line 
shows $B \propto \rho^{0.5}$. When the density becomes large, 
each relation approximately tends to $B \propto \rho^{0.5}$. 
In the case of $\beta_0=0.25$, the relation 
shows that core initially evolves to greater density without increasing 
the magnetic field strength. This is caused by the slip of neutral 
gas through the magnetic field during the subcritical phase of evolution.

\section{Conclusions and Discussion}
\label{conclusion}

We have studied fragmentation of a sheet-like self-gravitating cloud
by three-dimensional MHD simulations. The main results are as follows.

\begin{itemize}
\item
We confirmed that in the case of an initially 
subcritical cloud ($\beta_0=0.25$), cores  
developed gradually over an ambipolar diffusion time,
while the cores in an initially supercritical cloud
($\beta_0=4$ or $\beta_0=100$) developed
in a dynamical time. 

\item
The infall speed on to cores is subsonic in the case of an initially
subcritical cloud, while there is extended supersonic
infall in the case of an initially supercritical cloud.
This is consistent with the result of the two-dimensional
simulations of \citet{bas04}. In our three-dimensional simulations, 
we also find that the $z$-component of the velocity follows 
the same pattern.

\item
The size of cores for mildly supercritical cloud ($\beta_0=4$) 
is bigger than that of highly supercritical cloud ($\beta_0=100$). 
The size becomes smaller again when the magnetic field is stronger 
than the critical ($\beta_0=0.25$). This result is consistent with 
the two-dimensional linear analysis of \citet{cio06}.

\item
When the cloud is initially subcritical ($\beta_0=0.25$), 
the plasma $\beta$ in cores is greater than in the surroundings. 
In contrast to this, the plasma $\beta$ in cores is slightly lower 
than the surroundings when the cloud is initially supercritical
($\beta_0=4$).
The latter result is probably
caused by the modestly nonequilibrium state along $z$ during 
the evolution.

\item
In the $B-\rho$ plane, the snapshot of the relation between 
magnetic field strength ($B$) and density ($\rho$)
at different spatial points of the cloud overlaps with 
the evolutionary track of an individual core. 
When the density becomes large, each relation 
approximately tends to $B \propto \rho^{0.5}$. 
\end{itemize}

Our simulation is the first fully three-dimensional simulation 
to study the role of magnetic fields and ion-neutral friction 
in fragmentation. In this paper, we concentrated
on the effect of initially small perturbations,
partly as a way to compare with established predictions of 
linear theory. 
Our models also serve as a guide to understand fragmentation 
occurring exclusively in dense subregions of clouds that 
contain only subsonic or transonic motions.
Real molecular clouds certainly 
contain supersonic turbulence, at least in their low-density
envelopes, as is observed through large line-widths of 
emission lines from relatively low-density tracers. 
The additional effect of supersonic turbulence on 
three-dimensional fragmentation
with magnetic fields and ion-neutral friction 
will be studied in an upcoming paper.

\section*{Acknowledgments}
SB was supported by a research grant from NSERC.
The numerical computations were done mainly on the VPP5000
at the National Astronomical Observatory of Japan.

\end{document}